\documentclass[aps,prd,superscriptaddress]{revtex4}
\newcommand{\ra}{\rightarrow}

\newcommand{\be}{\begin{equation}}
\newcommand{\ee}{\end{equation}}
\newcommand{\bea}{\begin{eqnarray}}
\newcommand{\eea}{\end{eqnarray}}

\newcommand{\s}{\smallskip}

\newcommand{\bc}{\begin{center}}
\newcommand{\ec}{\end{center}}
\newcommand{\bu}{\begin{underline}}
\newcommand{\eu}{\end{underline}}

\newcommand{\ty}{\textstyle}
\newcommand{\ifp}{inverse Faddeev-Popov operator}
\begin{document}
\draft
\title{The Coulomb interaction and the inverse Faddeev-Popov operator in QCD}
\author{Kurt Haller\thanks{E-mail: Kurt.Haller@uconn.edu}}
\affiliation{Department of Physics, University of Connecticut, Storrs, Connecticut
06269-3046} 
\author{Hai-cang Ren\thanks{E-mail}}
\affiliation{Physics Department, The Rockefeller University, 1230 York Avenue, New York, NY 10021-6399}

\date{\today}
\begin{abstract}
We give a proof of a local relation between the inverse Faddeev-Popov operator and 
the non-Abelian  Coulomb interaction between
color charges. 
\end{abstract}
\maketitle

\narrowtext

When the QCD Hamiltonian is expressed entirely in terms of gauge-invariant variables, a nonlocal
operator, $\Gamma^{\,ab}({\bf y},{\bf x})$, appears in the role of 
the non-Abelian analog
of $(8\pi|{\bf y}-{\bf x}|)^{-1}$, the `static' interaction between electric charges in Coulomb-gauge
QED. $\Gamma^{\,ab}({\bf y},{\bf x})$ has the form 
\be
\Gamma^{\,ab}({\bf y},{\bf x})=-{\ty \frac{1}{2}}{\cal C}^{ab}({\bf y},{\bf x})
\label{eq:GC}
\ee
with
\be
{\cal C}^{ab}({\bf y},{\bf x})={\int}d{\bf r}{\cal D}^{aq}({\bf y},{\bf r})\partial^2_{({\bf r})}
{\cal D}^{qb}({\bf r},{\bf x})\,,
\label{eq:ci1}
\ee
where ${\cal D}^{ab}({\bf y},{\bf x})$ is the \ifp.  
Evaluating the \ifp~${\cal D}^{ab}({\bf y},{\bf x})$, by calculating
its expectation value in a particular state vector or by some other means, is important for determining
the boundaries of the regions within which it is bounded.~\cite{gribov,zwanz1,zwanz2} Evaluating 
$\Gamma^{\,ab}({\bf y},{\bf x})$ is necessary for calculating the forces between colored objects,
such as those between heavy static quarks. The conjecture that the unboundedness of 
${\cal D}^{ab}({\bf y},{\bf x})$ as $|{\bf y}-{\bf x}|{\ra}\infty$ is related to the unbounded growth of the
force between color-bearing objects, and thereby to color-confinement,~\cite{gribov,zwanz1,zwanz2} suggests a 
close relationship between points at which ${\cal D}^{ab}({\bf y},{\bf x})$ and 
${\Gamma}^{ab}({\bf y},{\bf x})$ become unbounded. \s

In addition to the {\em nonlocal} relation expressed in Eq.~(\ref{eq:ci1}), there is also a {\em local} 
 relation between $\Gamma^{\,ab}({\bf y},{\bf x})$ and ${\cal D}^{ab}({\bf y},{\bf x})$, 
\be
{\cal C}^{ab}({\bf y},{\bf x})=\frac{{\partial}\left(g{\cal D}^{\,ab}({\bf y},{\bf x})\right)}
{{\partial}g}\,,
\label{eq:cdrel}
\ee
which, to the best of our knowledge, first appeared in a paper by Swift.~\cite{swift} Because 
Eq.~(\ref{eq:cdrel}) expresses $\Gamma^{\,ab}({\bf y},{\bf x})$ as a local functional of 
${\cal D}^{ab}({\bf y},{\bf x})$, it makes the relation between the infrared behavior of
$\Gamma^{ab}({\bf y},{\bf x})$ and that  of ${\cal D}^{ab}({\bf y},{\bf x})$ 
much more transparent.\s

The inverse Faddeev-Popov operator ${\cal D}^{\,ab}({\bf y},{\bf x})$ is defined by the relation
\be
{\partial}{\cdot}D^{ca}_{({\bf y})}\,{\cal D}^{\,ab}({\bf y},{\bf x})=\delta_{cb}\delta({\bf y}-{\bf x})
\label{eq:FPd} 
\ee
where
\be
\partial{\cdot}D_{{({\bf x})}}^{ab}=\frac{\partial}{{\partial}x_i}
\left(\frac{\partial}{{\partial}x_i}\delta_{ab}
+gf^{a{q}b}A^{q}_i({\bf x})\right)\,;
\label{eq:tcom}
\ee
$A^{q}_i({\bf x})$ represents a transverse gauge field. It can, for example, be the gauge field in the 
Coulomb gauge; or it might be the gauge-invariant field $A^q_{{\sf GI}\,i}$ constructed within the Weyl
($A_0=0$) gauge,~\cite{CBH2} which has been identified with the Coulomb-gauge field.~\cite{khren} 
 ${\cal D}^{\,ab}({\bf
y},{\bf x})$ can be represented as the series
\be
{\cal D}^{\,ab}({\bf y},{\bf x})=\sum_{n=0}^\infty{\cal D}_{(n)}^{\,ab}({\bf y},{\bf x})
\label{eq:Dsum}
\ee
with 
\bea
{\cal D}_{(n)}^{\,ab}({\bf y},{\bf x})&&=g^n{f}^{\vec{\alpha}ab}_{(n)}
{\int}\frac{d{\bf z}({\scriptstyle 1})}{4{\pi}|{\bf y}-
{\bf z}({\scriptstyle 1})|}A_{l_1}^{{\alpha}_1}
({\bf z}({\scriptstyle 1}))\frac{\partial}{{\partial}z({\scriptstyle 1})_{l_1}}
{\int}\frac{d{\bf z}({\scriptstyle 2})}
{4{\pi}|{\bf z}({\scriptstyle 1})-{\bf z}({\scriptstyle 2})|}
{\times}\nonumber \\
&&\!\!\!\!\!\!\!\!\!A_{l_2}^{{\alpha}_2}
({\bf z}({\scriptstyle 2}))\frac{\partial}{{\partial}z({\scriptstyle 2})_{l_2}}\;
{\cdots}{\int}\frac{d{\bf z}({\scriptstyle n})}{4{\pi}|{\bf z}({\scriptstyle n-1})-
{\bf z}({\scriptstyle n})|}
A_{l_n}^{{\alpha}_n}
({\bf z}({\scriptstyle n}))\frac{\partial}{{\partial}z({\scriptstyle n})_{l_n}}
\frac{1}{{4{\pi}|{\bf z}({\scriptstyle n})-{\bf x}|}}\,;
\label{eq:Pihermg}
\eea   
${f}^{\vec{\alpha}ab}_{(n)}$ represents the chain of SU(N) structure constants
 \be
{f}^{\vec{{\alpha}}bh}_{(n)}={f}^{{\alpha}_1bu_1}\,\,{f}^{u_1{\alpha}_2u_2}\,
{f}^{u_2{\alpha}_3u_3}\,\cdots\,
\,{f}^{u_{(n-2)}{\alpha}_{(n-1)}u_{(n-1)}}{f}^{u_{(n-1)}{\alpha}_nh}\,,
\label{eq:fproductN}
\ee
where repeated superscripted indices are summed; the chain reduces for 
$n=1$ to ${f}^{\vec{{\alpha}}bh}_{(1)}={f}^{{{\alpha}}bh}$;
and for $n =0$, to ${f}^{\vec{{\alpha}}bh}_{(0)}=-\delta_{bh}$. 
These properties of ${f}^{\vec{\alpha}ab}_{(n)}$ enable us to conclude that,
for $n=0$ and $n=1$, the respective expressions for ${\cal D}_{(n)}^{\,ab}({\bf y},{\bf x})$ are
\be
{\cal D}_{(0)}^{\,ab}({\bf y},{\bf x})=\frac{-\delta_{ab}}{4{\pi}|{\bf y}-{\bf x}|}
\label{eq:D0}
\ee
and
\be
{\cal D}_{(1)}^{\,ab}({\bf y},{\bf x})=gf^{{\delta}ab}
{\int}\frac{d{\bf z}}{4{\pi}|{\bf y}-
{\bf z}|}A_{k}^{{\delta}}
({\bf z})\frac{\partial}{{\partial}z_k}\left(\frac{1}{4{\pi}|{\bf z}-{\bf x}|}\right)\,.
\label{eq:Da}
\ee
In Ref.~\cite{khren}, we pointed out that ${\cal D}^{\,ab}({\bf y},{\bf x})$ obeys
the integral equation~\cite{fs}
\be
{\cal D}^{\,ab}({\bf y},{\bf x})=-\left(\frac{\delta_{ab}}{4{\pi}|{\bf y}-{\bf x}|}+gf^{{\delta}au}
{\int}\frac{d{\bf z}}{4{\pi}|{\bf y}-{\bf z}|}A_{k}^{{\delta}}
({\bf z})\frac{\partial}{{\partial}z_k}{\cal D}^{\,ub}({\bf z},{\bf x})\right)\,.
\label{eq:PIe}
\ee
Eq.~(\ref{eq:PIe}) can also be obtained from the defining equation for 
${\cal D}^{\,ab}({\bf y},{\bf x})$,
\be
\left(\delta_{ah}{\partial}_{({\bf z})}^2+
gf^{a{q}h}A^{q}_{i}({\bf z}){\partial}_i^{({\bf z})}\right)
{\cal D}^{hb}({\bf z},{\bf x})=\delta_{ab}\delta({\bf z}-{\bf x})
\ee
by integrating both sides of the equation, as shown by
\be
-{\int}\frac{d{\bf z}}{4{\pi}|{\bf y}-{\bf z}|}\left\{\left(\delta_{ah}{\partial}_{({\bf z})}^2+
gf^{a{q}h}A^{q}_{i}({\bf z}){\partial}_i^{({\bf z})}\right)
{\cal D}^{hb}({\bf z},{\bf x})\right\}=\frac{-\,\delta_{ab}}{4{\pi}|{\bf y}-{\bf x}|}\,.
\ee
As was pointed out in Ref.~\cite{swift}, it is possible to represent ${\cal C}^{ab}({\bf y},{\bf x})$
as the series
\be
{\cal C}^{ab}({\bf y},{\bf x})=\sum_{n=0}^\infty{\cal C}^{ab}_{(n)}({\bf y},{\bf x})
\label{eq:cnser}
\ee
and to observe, from iterating Eq.~(\ref{eq:PIe}), 
that, order by order, each order examined confirms the relation 
\be
{\cal C}^{ab}_{(n)}({\bf y},{\bf x})=\frac{d}{dg}\,\left(g{\cal D}^{ab}_{(n)}({\bf y},{\bf x})\right)\,.
\ee
Ref.~\cite{swift} then points out that this fact can be used to prove Eq.~(\ref{eq:cdrel}). 
We will give a complete proof of Eq.~(\ref{eq:cdrel}) that does not require a perturbative
decomposition of 
${\cal C}^{ab}({\bf y},{\bf x})$. We use Eq.~({\ref{eq:PIe}}) to represent ${\cal D}^{\,bh}({\bf y},{\bf r})$, 
multiply both
sides of that equation by $\partial^2_{({\bf r})} {\cal D}^{qb}({\bf r},{\bf x})$, and integrate over ${\bf r}$, to obtain
\be
{\cal C}^{ab}({\bf y},{\bf x})={\cal D}^{ab}({\bf y},{\bf x})-
gf^{{\delta}au}{\int}\frac{d{\bf z}}{4\pi|{\bf y}-{\bf z}|}
A_{{\sf GI}\,k}^{\delta}({\bf{z}})\frac{\partial}{{\partial}z_k}{\cal C}^{ub}({\bf z},{\bf x})\,.
\label{eq:ci2}
\ee
We then {\em define} ${\bar {\sf C}}^{ab}({\bf y},{\bf x})$ as
\be
{\bar {\sf C}}^{ab}({\bf y},{\bf x})=\frac{{\partial}\left(g{\cal D}^{\,ab}({\bf y},{\bf x})\right)}
{{\partial}g}\,.
\ee 
and apply the operation of multiplying by $g$ and then differentiating with respect to $g$ to both sides
of Eq.~({\ref{eq:PIe}}), obtaining
\be
{\bar {\sf C}}^{ab}({\bf y},{\bf x})=-\frac{\delta_{ab}}{4{\pi}|{\bf y}-{\bf x}|}
\overbrace{-f^{{\delta}au}
{\int}\frac{d{\bf z}}{4{\pi}|{\bf y}-{\bf z}|}A_{k}^{{\delta}}
({\bf z})\frac{\partial}{{\partial}z_k}\sum_{n=0}^\infty(n+2)g^{n+1}{\sf d}_{(n)}^{\,ub}({\bf z},{\bf x})}
^{\sf R_2}\,,
\label{eq:PIe2}
\ee
where we use Eq.~(\ref{eq:Pihermg}) to write 
${\cal D}_{(n)}^{\,ab}({\bf y},{\bf x})=g^n{\sf d}_{(n)}^{\,ab}({\bf y},{\bf x})$ and 
where ${\sf d}_{(n)}^{\,ab}({\bf y},{\bf x})$ is independent of $g$.
We write the second term on the right-hand side of Eq.~(\ref{eq:PIe2})
\be
{\sf R_2}={\sf R_2}(A)+{\sf R_2}(B)
\ee
with 
\be
{\sf R_2}(A)=-gf^{{\delta}au}
{\int}\frac{d{\bf z}}{4{\pi}|{\bf y}-{\bf z}|}A_{k}^{{\delta}}
({\bf z})\frac{\partial}{{\partial}z_k}\sum_{n=0}^{\infty}g^{n}{\sf d}_{(n)}^{\,ub}({\bf z},{\bf x})=
-gf^{{\delta}au}
{\int}\frac{d{\bf z}}{4{\pi}|{\bf y}-{\bf z}|}A_{k}^{{\delta}}
({\bf z})\frac{\partial}{{\partial}z_k}{\cal D}^{\,ub}({\bf z},{\bf x})
\ee
and
\be
{\sf R_2}(B)=-gf^{{\delta}au}
{\int}\frac{d{\bf z}}{4{\pi}|{\bf y}-{\bf z}|}A_{k}^{{\delta}}
({\bf z})\frac{\partial}{{\partial}z_k}\sum_{n=0}^\infty(n+1)g^{n}{\sf d}_{(n)}^{\,ub}({\bf z},{\bf x})
=-gf^{{\delta}au}
{\int}\frac{d{\bf z}}{4{\pi}|{\bf y}-{\bf z}|}A_{k}^{{\delta}}
({\bf z})\frac{\partial}{{\partial}z_k}{\bar {\sf C}}^{ub}({\bf z},{\bf x})\,.
\ee
Since
\be
-\frac{\delta_{ab}}{4{\pi}|{\bf y}-{\bf x}|}+{\sf R_2}(A)={\cal D}^{\,ab}({\bf y},{\bf x}),
\ee
 it follows that
\be
{\bar {\sf C}}^{ab}({\bf y},{\bf x})={\cal D}^{\,ab}({\bf y},{\bf x})
-gf^{{\delta}au}{\int}\frac{d{\bf z}}{4\pi|{\bf y}-{\bf z}|}
A_{{\sf GI}\,k}^{\delta}({\bf{z}})\frac{\partial}{{\partial}z_k}{\bar {\sf C}}^{ub}({\bf z},{\bf
x})\,.
\label{eq:PIe3}
\ee
Since Eqs.~(\ref{eq:ci2}) and (\ref{eq:PIe3}) are identical, and both are linear integral
equations, ${\cal C}^{ab}({\bf y},{\bf x})$ and ${\bar {\sf C}}^{ab}({\bf y},{\bf x})$ are
identical as well, and Eq.~(\ref{eq:cdrel}) is proven. We note, also, that the fact that 
${\cal D}^{ab}({\bf y},{\bf x})={\cal D}^{ba}({\bf x},{\bf y})$,~\cite{khren} implies that
${\Gamma}^{ab}({\bf y},{\bf x})={\Gamma}^{ba}({\bf x},{\bf y})$.\s

An interesting consequence of this theorem is the proper generalization, to non-Abelian 
gauge theories,  of the static 
potential between charges in Abelian, Coulomb-gauge QED,
\be
{\int}d{\bf x}d{\bf y}\rho({\bf x})\frac{1}{8{\pi}|{\bf y}-{\bf x}|}\rho({\bf y})\equiv
-\frac{1}{2}{\int}d{\bf x}\rho({\bf x})\left(\frac{1}{\partial^2}\right)\rho({\bf x})\,.
\label{eq:qed}
\ee
We might, perhaps, wonder whether one could extend Eq.~(\ref{eq:qed}) to non-Abelian 
theories by replacing the Laplacian operator in Eq.~(\ref{eq:qed}) with the Faddeev-Popov operator
$\partial{\cdot}D$. But  Eq.~(\ref{eq:cdrel}) informs us that this `naive' substitution is not allowed.
The proper extension of Eq.~(\ref{eq:qed}) into the non-Abelian domain is to write the non-Abelian
nonlocal interaction between color-charges symbolically as
\be
\int\!d{\bf x}\left(j_0^a({\bf x})+
J_{0\,({\sf GI})}^{a\,{\sf T}\,\dagger}({\bf x})\right)
\left\{\frac{\partial\{g[(\partial{\cdot}D)^{-1}]^{ab}\}}{\partial g}\right\}
\left(j_0^b({\bf x})+
J_{0\,({\sf GI})}^{b\,{\sf T}}({\bf x})\right)
\label{eq:colb3}
\ee
where ${\partial{\cdot}D}$ is given by Eq.~(\ref{eq:tcom}).\s

Eq.~(\ref{eq:cdrel}) has significant advantages over Eq.~(\ref{eq:ci1}). 
For a fixed set of points ${\bf y}$ and ${\bf x}$, Eq.~(\ref{eq:ci1}) expresses
$\Gamma$ as a {\it{nonlocal}} functional of ${\cal D}$,  so that it is not very intuitive 
that the behavior of $\Gamma^{\,ab}({\bf y},{\bf x})$ as $|{\bf y}-{\bf x}|{\ra}\infty$ is related to 
the behavior of ${\cal D}^{ab}({\bf y},{\bf x})$ as $|{\bf y}-{\bf x}|{\ra}\infty$. 
In contrast, Eq.~(\ref{eq:cdrel})
expresses $\Gamma$ as a {\it{local}} functional 
of ${\cal D}$ and the relation between the infrared behavior of $\Gamma^{ab}({\bf y},{\bf x})$ and that 
of ${\cal D}^{ab}({\bf y},{\bf x})$ becomes more transparent. Moreover, as illustrated in the work
of Szczepaniak and Swanson,~\cite{szsw}, Eq.~(\ref{eq:cdrel}) enables 
one to eliminate an integration over one spatial variable in evaluating expectation values of the Hamiltonian
for trial wave functions that represent the physical QCD vacuum.\s

The authors thank Prof. E. S. Swanson for calling their attention to Refs.~\cite{swift} and \cite{szsw}
after their initial preprint had been posted. 
The research of KH was supported by the Department
of Energy under Grant No.~DE-FG02-92ER40716.00 and that of HCR was supported by the Department of Energy under Grant
No.~DE-FG02-91ER40651-TASKB.

\end{document}